\documentclass[11pt,a4paper]{article}

\usepackage[margin=1.5in]{geometry}

\usepackage{amssymb}
\usepackage{color}
\usepackage{sectsty}
\usepackage{times}
\usepackage{latexsym}
\usepackage{natbib}
\usepackage{longtable}
\usepackage{enumerate}
\usepackage{adjustbox}
\usepackage{graphicx}
\usepackage[super]{nth}
\usepackage{epsfig}
\usepackage{amsmath}
\usepackage{amssymb}
\usepackage{amsfonts}
\usepackage[english]{babel}
\usepackage{graphicx}
\usepackage{graphics}
\usepackage{psfrag}
\usepackage{amsbsy}
\usepackage{textcomp}
\usepackage{outlines}
\usepackage{grffile}
\usepackage[justification=centering]{caption}
\usepackage{authblk}

\def\bx{\boldsymbol{x}}

\def\b0{\boldsymbol{0}}
\def\b1{\boldsymbol{1}}

\def\bvarphi{\mbox{\boldmath $\varphi$}}

\title{Measurement error induced by locational uncertainty when estimating discrete choice models with a distance as a regressor}

\author[1]{Giuseppe Arbia}
\author[2]{Paolo Berta \thanks{Corresponding to: paolo.berta@unimib.it}}
\author[3]{Carrie B. Dolan}

\affil[1]{\small{Department of Statistics, Catholic University of Rome (Italy)}}
\affil[2]{\small{Department of Statistics and Quantitative Methods, University of Milano Bicocca (ITA)}}
\affil[3]{\small{Department of Kinesiology and Health Sciences, College of William and Mary (USA)}}

\begin{document}

\maketitle

\section{Introduction}
\label{sec:Introduction}

Spatial microeconometric studies typically suffer from several types of inaccuracies that are not present when dealing with the classical regional spatial econometrics models. Among those, missing data, locational errors, sampling without a formal sample design, measurement errors and misalignment are the typical sources of inaccuracy that can affect the results in a spatial microeconometric analysis. \\
\cite{arbia2015measurement} analysed intentional and unintentional measurement errors associated with the geo–masking of spatial micro–data. In some empirical circumstances, the exact geographical position of the individuals can be uncertain due to lack of information. This happens, for instance, when we have a list of firms in a small area (like e.g. a census tract), but we don’t know their exact address within the area. A typical case is when we use GPS position of individuals derived from cell phone information where the position is known depending on the corresponding accuracy of the phone GPS. In this case, it is common to assign the individual to the centroid of each area, but this procedure obviously generates a locational error. \cite{arbia2015measurement} referred to this situation as to \textit{unintentional locational error}. \citep{deardon2012spatial} found that when modelling the spread of disease at the individual level unintentional locational error can be accounted for through a random effects model. However, the biasing effect on the basic reproductive number is not easily alleviated. In other empirical cases, the individual’s position is perfectly known, but they are geo–masked a priori before being made it publicly available to the analysts, in order to preserve confidentiality. In this second case we say that an intentional positional error was introduced. In their paper, \cite{arbia2015measurement} examined the instability of the results induced by locational errors in spatial regression models that include a distance as a predictor. \cite{colenutt1968building} demonstrated how error can accumulate so that accurate prediction becomes challenging. \\
In economics and social science, several papers use spatial econometrics models including distance as regressor. In healthcare the adoption of econometrics techniques in the study of the patients choices produced a large number of papers in which the distance between patients and hospitals is one of the most important predictors. \cite{perucca2019spatial} used spatial error models to examine spatial inequalities in access to care using distance as a key measure of patient mobility. Both unintentional locational errors (that is inaccuracy in the data collection due to approximate address or lack of sufficient information when individuals are located at the centroid of a small area) and intentional locational errors (a-posteriori geo–masking of the exact GPS coordinates to preserve confidentiality), are potential sources of errors that may undermine the results and mislead the substantive conclusions.  \\
In this area we observe an increasing awareness and a growing literature aimed to integrate information on the precision of geographic data to improve the accuracy of modeling effort \citep{aerts2003accounting}. Several solutions to this problem have been proposed ranging from the efficient estimation of probability density to sensitivity models \citep{lilburne2009sensitivity} and, more recently, to the method termed geoSIMEX, which expands on the previous methods by integrating an extrapolation step into traditional Monte Carlo approaches in order to avoid bias due to spatial imprecision \citep{geosimex}. This approach certainly mitigates a key limitation of using simulation alone: namely the failure to account for spatial imprecision during the estimation of standard errors and parameter estimates during each iteration of a simulation, thus outperforming model averaging in its ability to capture the true coefficient at all levels of spatial imprecision. Excluding or making assumptions about spatially imprecise data can lead to biased estimated coefficients, potentially resulting in misleading policy conclusions. GeoSIMEX provides an additional step in communicating to the practitioners the resulting level of attenuation they should expect from a discrete choice regression analysis. This approach, however, lessens the risk of committing a type I error, but it does heighten the risk of committing a type II error. \\
As a matter of fact, further work is needed to refine uncertainty procedures and lessen the impact of spatial uncertainty on the estimation of policy relationships and this paper wants to contribute to this important growing literature.\\
In the paper already quoted, \cite{arbia2015measurement} examined the effects of intentional locational errors induced by geo–masking in the specific case of continuous linear regression models when a distance is used as a regressor. In this paper we aim to extend these results to the case of non-linear models. We aim to shed light on the distortion effects due to locational error to make researchers aware of the possible limitations of their inferential conclusions. In contrast with the case of a continuous linear regression model, in the case of a non-linear model, the ML estimators do not have closed form solutions since they are usually derived by numerical maximization. As a consequence, in this paper no formal result can be obtained in terms of bias and we have to resort to a Monte Carlo (MC) approach. Some formal results, however, can be obtained related to the loss in the efficiency of the estimates when individual spatial data are contaminated with locational errors. \\
The rest of the paper is structured as follows. In Section \ref{sec:sec2} we will summarize the main results of \cite{arbia2015measurement} and we will present a motivating case study based on \cite{berta2016association}. Section \ref{sec:sec3} will present the results of several Monte Carlo experiments on the bias induced on spatial model parameters by intentional or unintentional locational errors, whereas Section \ref{sec:sec4} describes their effects on the efficiency of the Maximum Likelihood (ML) estimators. Finally, Section \ref{sec:sec5} concludes the paper.

\section{Effects of geo-masking in the regression analysis of healthcare competition}
\label{sec:sec2}

The theoretical motivation of this paper refers to \cite{arbia2015measurement}, where the authors study the negative effects of the geo-masking, examining the measurement error introduced by geo-masking the individuals' true location, when distances are used as predictors in a linear regression. A very popular geo--masking mechanism is the uniform geo--masking (explained, e. g., in \citep{burgert2013geographic}), in which the true coordinates are transformed by displacing the individuals' position along a random angle (say $\theta^*$) and a random distance (say $\delta$) both following a uniform probability law. The mechanism can be formally expressed through the following hypotheses:

\noindent \textbf{HP1}: $\theta \sim U(0, \theta^*$) and $\delta \sim U (0^\circ,360^\circ)$, with $\theta^*$ the maximum distance error; \\
\textbf{HP2}: $\theta$  and  $\delta$ are independent. 

\begin{figure}
\centering
\includegraphics[width=0.9\textwidth]{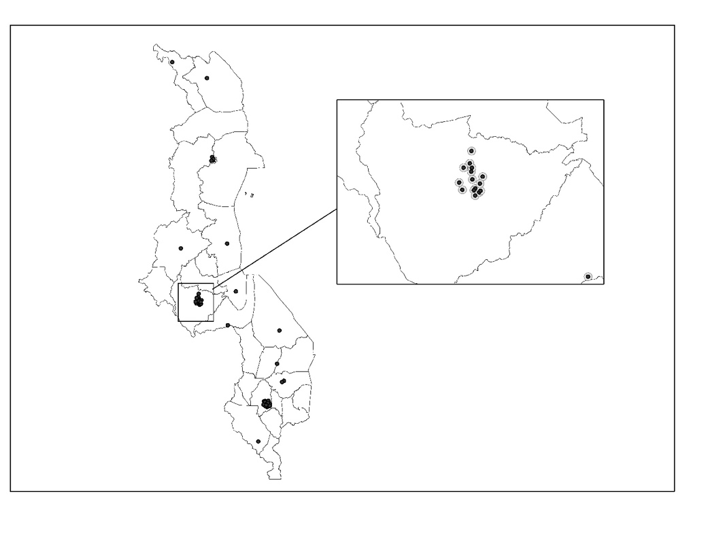}
\caption{Random Point Displacement, Urban Malawi DHS Clusters.}
\label{fig1}
\end{figure}

\noindent An example of uniformly geo--masked locations is illustrated in Figure \ref{fig1}. Represented as points in Figure \ref{fig1} are groupings of urban households that participated in the 2014 Malawi Malaria Indicator Survey \citep{malariamalawi}. The buffer represents the maximum amount of random displacement introduced by the Demographic Health Survey (DHS) to ensure that respondent confidentiality is maintained. The true location of the respondent households are located within this 2km buffer. 
\noindent An alternative geo--masking mechanism is the Gaussian geo--masking where points are randomly reallocated in the neighborhood of the true location, following a Gaussian bivariate density function with the mean vector coinciding with the true point and a given variance which can be expressed again as a function of the (practical, 99\%) maximum displacement distance $\theta^*$.\\
Let us now consider a simple linear regression model using a distance as a predictor:
\begin{equation}
y_{ih} = \alpha + \beta d^{2}_{ih} + \varepsilon_{ih}
\end{equation}
with $d_{ih}$ the distance between point $i$ and point $h$. Considering a healthcare framework, patients living in a point are moved (geo--masking their true coordinates) within a circle with a maximum radius of, say, $\theta^*$ and randomly re-assigned in an erroneous position. In this way, since the coordinates associated to the geo--masked point are considered when calculating the distance of the patient from an hospital, we introduce a measurement error in the independent variable. \cite{arbia2015measurement} extended the classical error measurement theory (e. g., \cite{verbeek2008guide}) to this specific case by showing that the greater the maximum displacement distance ($\theta^*$), the larger will be both the loss in efficiency and the bias of OLS estimator of the $\beta$ parameter, producing a reduction towards zero of its absolute value (known in the literature as the attenuation effect).\\
The loss in efficiency and the attenuation effect observed in the presence of geo--masking, are very important under a practical point of view. Figure 2 reports the theoretical behaviour of the attenuation effect for Gaussian and uniform geo-masking of points as a function of the maximum displacement distance $\theta^*$ \citep{arbia2015measurement}. The inspection of the graph clearly shows that the attenuation increases dramatically already at small levels of $\theta^*$, and the Gaussian geo--masking produces more severe consequences on the estimation of $\beta$ than the uniform geo-masking. 
\begin{figure}
\centering
\includegraphics[width=0.9\textwidth]{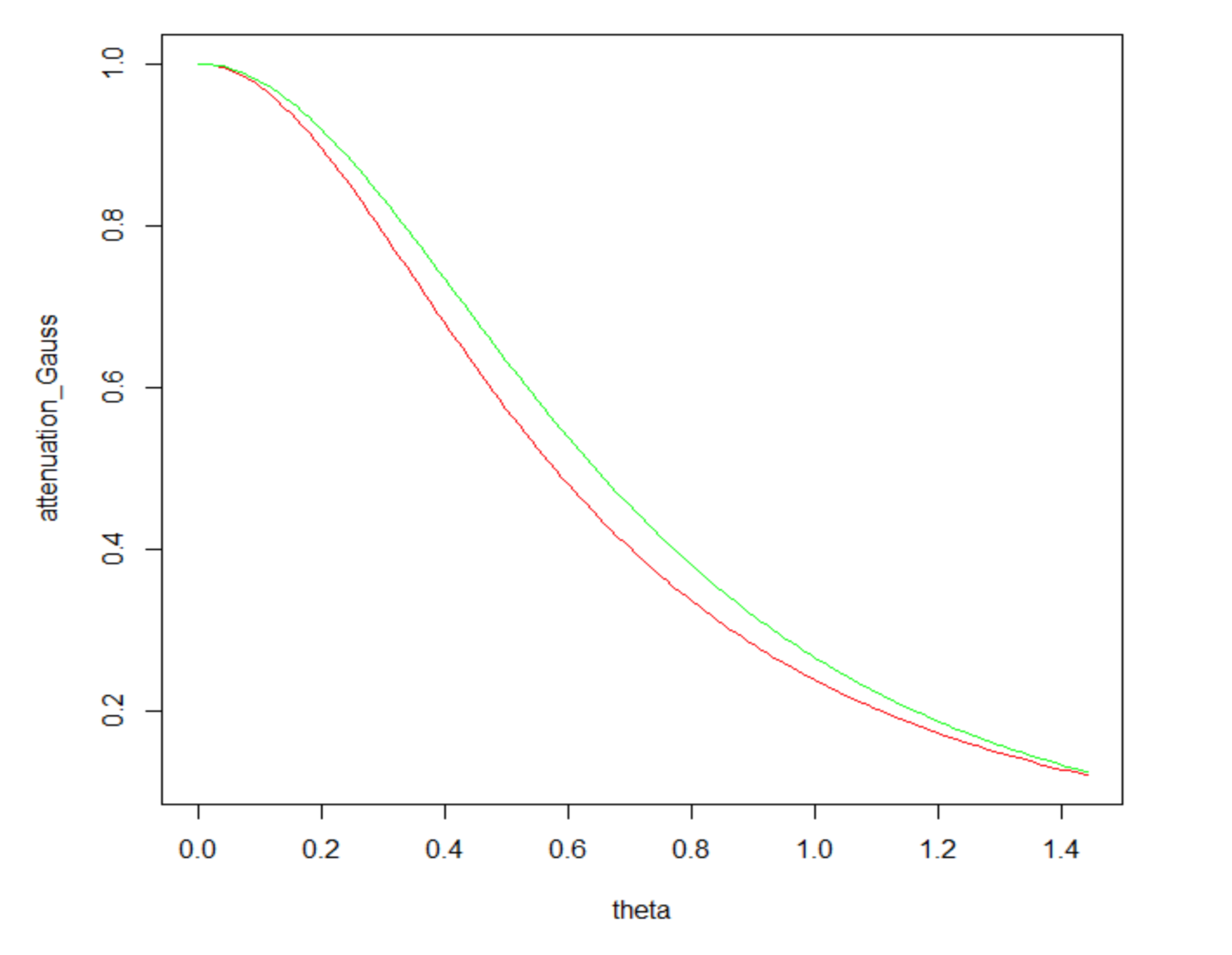}
\caption{Attenuation effect in the presence of geo--masking as a function of the maximum displacement distance $\theta^*$. True parameter value = 1. Gaussian geo--masking (red line). Uniform geo--masking (green line). Source: \cite{arbia2015measurement}.}
\label{fig2}
\end{figure}

\noindent The analysis of the competition in healthcare is one of the most widely studied topics in health economics and health econometrics. This research area follows the approach suggested by \cite{kessler2000hospital}. In this seminal paper, the authors analyse the relationship between hospital quality and competition, modelling the patients' choices in order to build a competition index (Herfindahl-Hirschman index). This measure of hospital competition based on the predicted patients' choices is included as a predictor in a model, and the hospital quality as the dependent variable. Following this approach, \cite{berta2016association} assumed that the discrete choice of the single patient $i$ of choosing hospital $h$ (say, $y_{ih}$) is related to the expected utility of the patient $y^{*}_{ih}$, with $y_{ih} = 1$ if $y^{*}_{ih} > 0$, and $0$ otherwise.\\
The utility model was specified as follows:
\begin{equation}
y^{*}_{ih} = \rho d_{ih} + \delta_{h} Network_{ih} + \gamma_{h} GP_{ih} + \xi_{h} \bx_{ih}
\end{equation}
where $d_{ih}$  is the distance between patient $i$ and hospital $h$ expressed in minutes of travelling, $GP_{ih}$ is the percentage of patients living in the same zip code as patient $i$ and sharing the GP with patient $i$ and admitted to the hospital $h$, while $\bx_{ih}$ is a set of patient-level characteristics. The variable $Network_{ih}$ is a continuous variable representing the share of people living in the same municipality as patient $i$ and admitted in the same hospital $h$ in the 12 months before the admission of patient $i$. Considering that the travel distance is strictly correlated with the hospital choice, the coefficients related to the distance was expected to be negative. The model was estimated using an administrative dataset related to 8,627 patients admitted in the 20 cardiac surgery wards located in Lombardy (Italy) in 2014, obtaining a total number of 172,540 observations. In \cite{berta2016association}, as in the majority of the papers adopting this empirical strategy, the patient location was not perfectly known, and it was approximated by the centroid of the municipality of the patient.  \\
The aim of this paper is to extend the results of \cite{arbia2015measurement} to the discrete choice models used in \cite{berta2016association} thus shedding light on the distorting effects of locational errors in this instance. However, in contrast with the case of a continuous linear regression model, in the case of a non-linear model, the ML estimators do not have closed form solutions and they are usually derived by numerical maximization. 

\section{Monte Carlo evaluation of locational errors in discrete choice models}
\label{sec:sec3}

\subsection{A first simulation experiment: the effect of ``unintentional'' locational error when allocating the individuals in the centroids of the areas}
\label{sub3.1}

\begin{figure}
\centering
\includegraphics[width=0.9\textwidth]{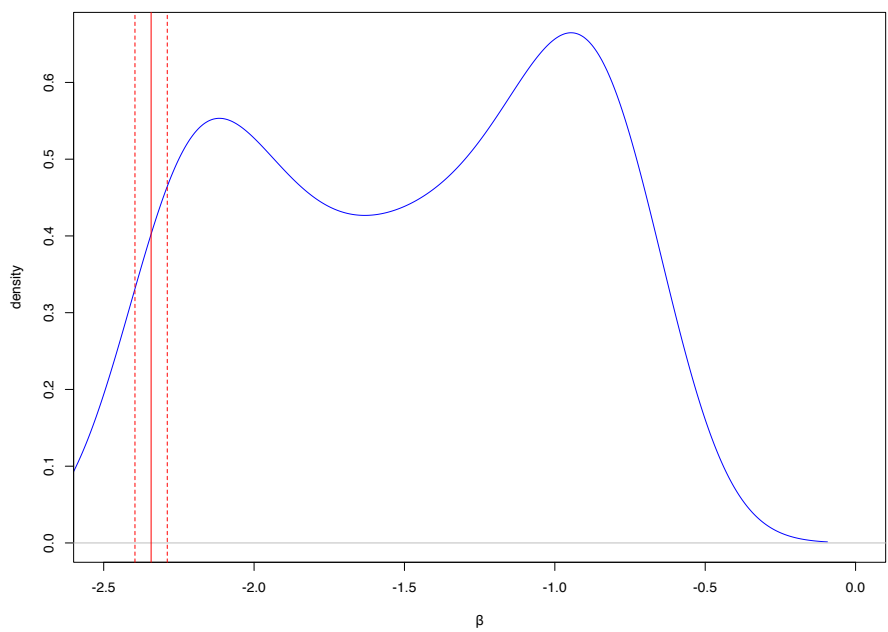}
\caption{MC distribution of the models' coefficients $\beta$ after geo--masking compared with the $\beta$ based on the ``true'' location (solid vertical line) and its confidence interval (dashed vertical lines).}
\label{fig4}
\end{figure}

In a first MC experiment, we aimed to quantify the existence of distortion effects in discrete model estimation, specifically in the case of unintentional locational error induced by uncertainty on individual's location. In this MC study, we considered the dataset used by \cite{berta2016association} for their healthcare competition model and we assumed that the individuals' location observed by the authors was the true patient location known without error. We then repeated the same logit regression analysis of the previously referenced paper randomly relocating 1,000 times the 8,627 patients of the dataset using a uniform geo--masking (see Section \ref{sec:sec2}) with a maximum distance $\theta^*$ which equals the radius of a circle with the equivalent surface area of each municipality. This radius represents the (approximate) maximum location error committed when an individual is allocated to the centroid. When a point is randomly relocated outside the study area the point it is randomly relocated a second time. The 1,000 simulated relocations of the patients thus define 1,000 new matrices of distances between the patients and the hospitals. Using these modified distances, we estimate 1,000 discrete choice models, where the dependent variable is the patients' choice and the covariate is the distance. In this way, we obtained 1,000 replications of the estimates of  the parameter $\beta$ concerning the effect of the distance on the patients' choice.\\
The simulation results are reported in Figure \ref{fig4}. Figure \ref{fig4} reports the kernel density for the MC distribution of the estimates of the parameter $\beta$ after geo--masking. The distribution assumes a symmetric shape. In Figure \ref{fig4} the kernel density of $\beta$ is compared with the $\beta$ parameter estimated with the not-distorted data (the straight red line). In addition, the two dashed red lines represent the confidence interval for the original $\beta$ parameter. Comparing the value obtained with the MC experiment we observed that only in the 10\% of the provided simulations do not statistically differ from the ``true'' $\beta$ \citep{berta2016association}, whereas in the 90\% the absolute value is lower and approaching 0.

\noindent This first simulation experiment clearly shows that a locational error originated by randomly allocating the patients within the specific municipality of reference, affects significantly the estimates of the parameters of a logit model. Indeed, our results reinforce those included in \cite{berta2016association}. In light of the effect reported here it is reasonable to believe that their results underestimate the spatial effect of distance in an hospital choice model.

\subsection{A second simulation experiment: the effects of ``intentional'' locational error with different displacement distances}
\label{sub3.2}

\begin{figure}
\centering
\includegraphics[width=0.9\textwidth]{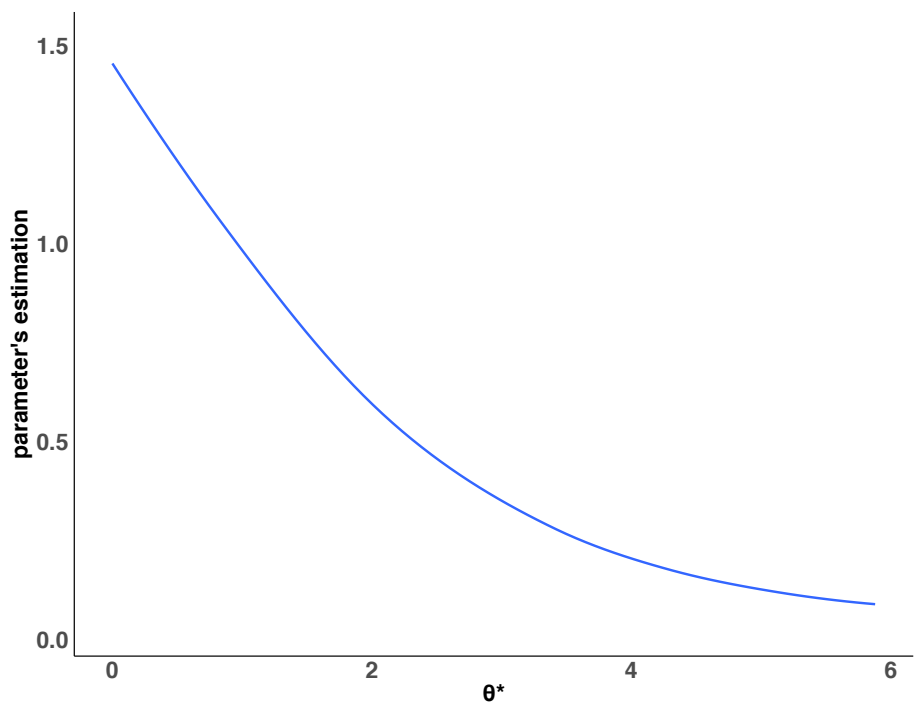}
\caption{Attenuation effect for 8,627 patients in Lombardy \citep{berta2016association} affected by uniform geo--masking for increasing displacement $\theta^*$. True parameter value = 1.5}
\label{fig5}
\end{figure}

In the first simulation exercise the maximum displacement distance was considered fixed and (since we imposed the constraint that a patient should remain in the original municipality) it depends only on the surface area of the municipality. The larger the municipality the larger the maximum locational error that can be introduced with the geo--masking mechanism. In this second simulation experiment, we aim at a more general result by removing the constraint related to the dimension of the municipality so as to be able to explore (similarly to the study in \cite{arbia2015measurement} for linear models) how the distortion in the estimation is related to the maximum radius of geo--masking. This second MC exercise, therefore, tends to reproduce the case of intentional locational error. In this second instance examined, for simplicity but without loss of generality, we considered the presence of only one hospital in the study area.
We thus consider the same set of individuals located in Lombardy and reported in the study by \cite{berta2016association}, and the associated Euclidean distances (say $d_{ih}$) measured between each patient $i$ and the hospital. In order to monitor the effects of geo--masking on the bias in the estimation of different displacement distances, we allowed the parameter $\theta^*$ to assume different values. In particular, we considered the following values: $\theta^* = 6.6 (0.1, 0.2, 0.3, 0.4, 0.5, 0.6, 0.7, 0.8, 0.9, 1)$, the constant 6.6 representing the radius of the equivalent circle with a surface area equal to that of the entire Lombardy region (after coordinates standardization). For each value of $\theta^*$, we then repeat 1,000 times a uniform geo--masking of the patients' coordinates. In this way, for each individual (and for each of the 10 maximum radius), we obtain 1,000 new distances (say $\bar{d}_{ih}$ ) and, for each of these geo--masked distances, we estimate again the patients' hospital choice. In particular, we assume that the observable choice of patient $i$ of being admitted to the hospital $h$ ($y_{ih}$) is related to the expected utility of $i$ choosing $h$ ($y^{*}_{ih}$), according to $y_{ih} = 1$ if $y^{*}_{ih} > 0$. As a consequence, our choice model is:
\begin{equation}
y_{ih} = \alpha + \beta \bar{d}_{ih}
\end{equation}
that can be estimated by adopting a logit model. We still expect the coefficient $\beta$ to be negative, but we also expect again that the geo--masking forces the coefficients towards 0 when the radius of displacement increases. Furthermore, similarly to the continuous linear model case, this distortion depends on the maximum displacement distance $\theta^*$ used.  The main results of this second simulation are reported in Figure \ref{fig5} which is the counter--part of Figure \ref{fig2} for the case considered in our study.
For increasing values of $\theta^*$ the coefficient related to the distance decreases monotonically towards 0, thus showing a similar effect to the attenuation effects observed in linear models \citep{arbia2015measurement}. A sharp decrease is observed already for low levels of $\theta^*$. For instance, when $\theta^{*} = 2$ (corresponding to 30\% of the maximum distortion distance) the estimated $\beta$ is about $0.6$ which is 40\% of the original value. Furthermore, the 92\% of the estimates provided in this simulation experiment differ from the original $\beta$. In terms of variability, the 42\% of the coefficients are not significant.

\subsection{Third simulation experiment: the effects of ``intentional'' locational error with 1,000 simulated individuals}
\label{sub3.3}

\begin{figure}
\centering
\includegraphics[width=0.9\textwidth]{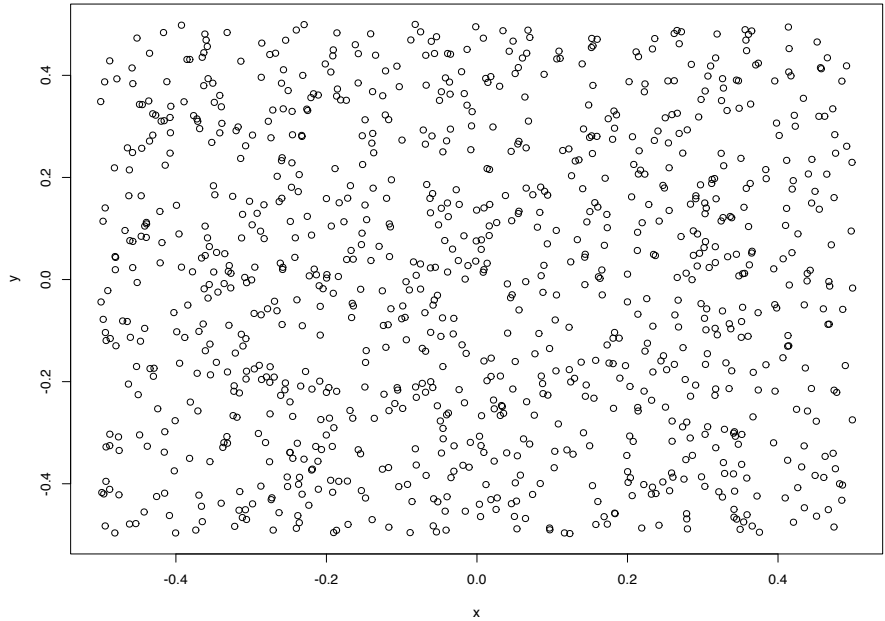}
\caption{1,000 individuals randomly distributed (CSR) in a squared unitary study area centred in the origin.}
\label{fig6}
\end{figure}

\begin{figure}
\centering
\includegraphics[width=0.9\textwidth]{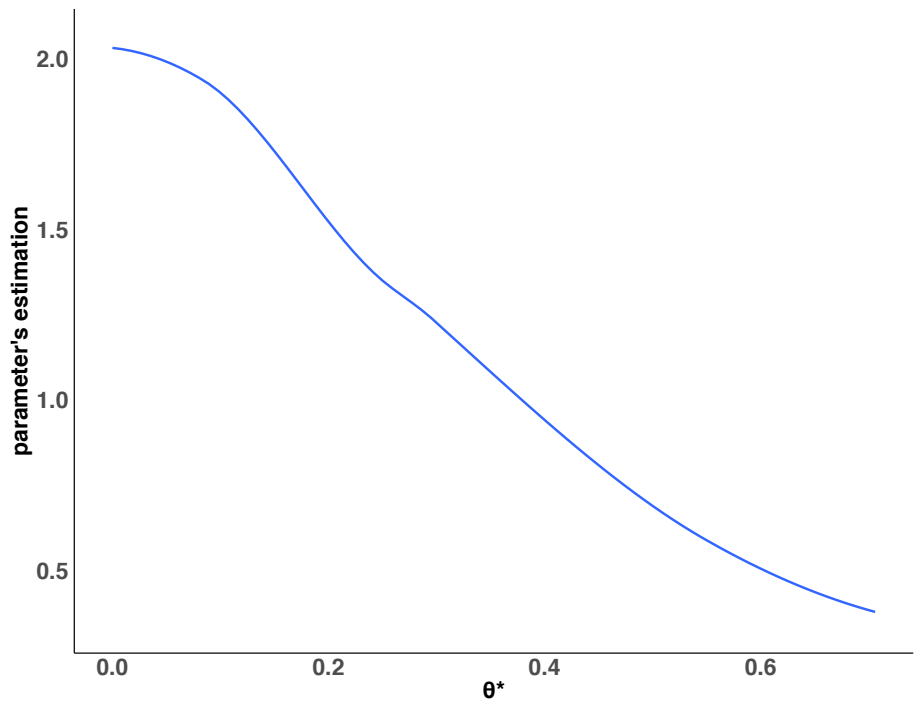}
\caption{Attenuation effect for 1,000 simulated individuals in the presence of uniform geo-masking as a function of the maximum displacement distance $\theta^*$. True parameter value $= 2$}
\label{fig7}
\end{figure}

To obtain more general results then those described in Subsection \ref{sub3.2}, in the third MC experiment we still refer to the case of an intentionally induced locational error, but we now abandon the reference to real data and we simulate the behaviour of $n = 1,000$ individuals randomly distributed in a unitary squared area centred on zero. Their distribution is reported in Figure \ref{fig6}.
Using the Complete Spatial Randomness (CSR) model \citep{diggle1983statistical} the individuals' coordinates are generated as two independent uniform distributions $U(-0.5, 0.5)$. For each individual located in $(i,j)$, we then calculate the Euclidean distance from the point $(0,0)$ assuming, without loss of generality, that the hospital is located in the centre of the study area. We then simulate a conditional binomial choice of the hypothetical hospital located at the origin $(0,0)$, obtaining a simulated $logit$ model, as follows: 
\begin{equation}
logit(\pi_{ij}|d_{ij},\bvarphi) = 1 - 2 * d_{ij}
\label{eq:4}
\end{equation}
where $logit(\pi_{ij}|d_{ij},\bvarphi)$  is the logit of the probability associated to a random conditional binomial distribution given the regression parameter vector $\bvarphi = (1, 2)$.\\
Starting from this set of data we fix again 10 maximum radius of distortions such that $\theta^* = 0.707 (0.1, 0.2, 0.3, 0.4, 0.5, 0.6, 0.7, 0.8, 0.9, 1)$, the constant $\sqrt{2}/2=0.707$ now representing half of the diagonal of the unitary square on which the data are laid.\\
For each of the 10 maximum $\theta^*$ we again replicate 1,000 times for each individual a uniform geo--masking mechanism. In this way, for each individual and for each of the 10 levels of the maximum radius, we obtain 1,000 new distances. Using the polar coordinates, the distorted distances from the origin ($\bar{d}_{ij}$) are calculated as follow:

\begin{equation}
\bar{d}_{ij} = \sqrt{(i-\theta \cos(\delta))^2+(j-\theta \sin(\delta))^2}
\end{equation}
where, as discussed in Section \ref{sec:sec2}, $\theta$ is the distance from the true point and $\delta$ the angle of the segment joining the true point with the displaced one. For each replication, we then re-estimate the logit model in equation \ref{eq:4} using the distorted distances instead of the true ones: 

\begin{equation}
logit(\pi_{ij}|\bar{d}_{ij},\bvarphi) = 1+2 * \bar{d}_{ij}
\label{eq:5}
\end{equation}
We expect again to observe an attenuation effect reducing $\hat{\beta}$ in absolute value towards 0 as the maximum radius of distortion increases. The mean of the estimates of $\hat{\beta}$ for each $\theta^*$ can be plotted over the increasing values of $\theta^*$, representing, once again, the attenuation effect on $\hat{\beta}$.
Results are reported in Figure \ref{fig7}.

As expected, for increasing values of $\theta^*$ the coefficient $\beta$ decreases towards 0 thus erroneously suggesting a not-significant relationship of the hospital choice based on the distance between the hospital and the place where the patient lives. Again, the decrease is very sharp. When only a small amount of distortion is imposed the value of $\beta$ is almost unaffected, but as soon as it increases further, we observe a sharp decline. In this case, only the 22\% of the estimates are different from the original $\beta$, and almost the 70\% is statistically equal to 0.

\section{The theoretical efficiency loss of ML estimates of the parameters of a logit model}
\label{sec:sec4}

As said in the introduction, the ML estimates of the parameters estimated in the previous experiments cannot be expressed in a closed form and thus there is no possibility to derive a formal relationship of the attenuation effect as it is done in the classical error measurement theory for linear regression models. However, it is possible to produce results in the case of examining efficiency loss due to geo--masking. In fact, let us consider the log-likelihood associated with the logit model in the case of perfect knowledge of the spatial coordinates. This can be expressed as:

\begin{equation}
\ell (\beta)  =  ln[L(\beta)] = \sum^{n}_{i=1} \left\lbrace y_i lnF(d_{ij} \beta) + (1-y_i) ln[1-F(d_{ij} \beta)] \right\rbrace 
\label{eq_10}
\end{equation}
where $F$ is a cumulative probability distribution function to be specified \citep{greene2016econometric}.\\
Now assume that the distance appearing in Equation \ref{eq_10} is affected by a measurement error which in turn is due to a locational error introduced intentionally by geo-masking to preserve the respondent confidentiality. Let us further consider the associated, error-affected, likelihood that can be expressed as:

\begin{equation}
 \bar{\ell} (\beta)  =  ln[\bar{L}(\beta)] = \sum^{n}_{i=1} \left\lbrace  y_i lnF(\bar{d}_{ij} \beta) + (1-y_i) ln[1-F(\bar{d}_{ij} \beta)] \right\rbrace 
\label{eq_11}
\end{equation}
where, as before, $\bar{d}_{i,j}$ represents the error--affected distance.\\
Let us now assume, in particular, that the cumulative probability distribution is specified as a standardized \textit{logistic} distribution, say $\Lambda$, characterized by 0 expected value and variance $\pi^{2}/3$. In this case, using Equation \ref{eq_10}, the associated score functions can be written as \citep{greene2016econometric}:

\begin{equation}
\frac{\partial}{\partial\beta}\ell (\beta)  =   \sum^{n}_{i=1} (y_i - \Lambda_i)d_{ij} = 0
\label{eq_12}
\end{equation}
with $\Lambda_i = \Lambda(d_{i,j},\beta)$, and the second order derivative as:

\begin{equation}
\frac{\partial^2}{\partial^2 \beta^2} \ell (\beta)  =   \sum^{n}_{i=1}  \Lambda_i(1 - \Lambda_i)d^{2}_{ij} = 0
\label{eq_13}
\end{equation}
with similar expressions for the first and the second likelihood derivatives, easy to obtain in the case of geo--masked coordinates. Using Equation \ref{eq_13}, we can obtain the elements of the Fisher Information Matrix related to the simulated true data as well as those related to the data after geo--masking.
Indeed, from Equation \ref{eq_13} we have: 
\begin{equation}
Var(\hat{\beta}) = -E \left[ \frac{\partial^2}{\partial^2 \beta^2} \ell (\beta) \right]  =  -E  \left[  \sum^{n}_{i=1}  \Lambda_i(1 - \Lambda_i)d^{2}_{ij} \right] 
\label{eq_14}
\end{equation}
and similarly, for the likelihood under the geo--masked coordinates:
\begin{equation}
Var(\hat{\bar{\beta}}) = -E \left[ \frac{\partial^2}{\partial^2 \beta^2} \ell (\beta) \right]  =  -E  \left[  \sum^{n}_{i=1}  \Lambda_i(1 - \Lambda_i)\bar{d}^{2}_{ij} \right] 
\label{eq_15}
\end{equation}
Equation \ref{eq_14} and \ref{eq_15} show that the variance of the estimators depends essentially on the distance $d$. If this distance is inflated by geo--masking the precision of the estimators will be reduced.\\
To show more explicitly this Efficiency Loss (EL), making use of Equations \ref{eq_14} and \ref{eq_15}, we can calculate the loss in the efficiency of the ML estimator as the ratio as the variance of the MLE with the geo--masked coordinates, say $\hat{\bar{\beta}}$, and the variance of the MLE with the true coordinates, say $\hat{\beta}$ :

\begin{equation}
EL = \frac{Var(\hat{\beta})}{Var(\hat{\bar{\beta}})} = \frac{ -E  \left[  \sum^{n}_{i=1}  \Lambda_i(1 - \Lambda_i)d^{2}_{ij} \right] }{ -E  \left[  \sum^{n}_{i=1}  \Lambda_i(1 - \Lambda_i)\bar{d}^{2}_{ij} \right] }
\label{eq_19}
\end{equation}
Now let us concentrate our attention on the denominator of this expression, where we have:
\begin{equation}
Var(\hat{\bar{\beta}}) =  E  \left[  \sum^{n}_{i=1}  \Lambda_i(1 - \Lambda_i)\bar{d}^{2}_{ij} \right] =  \sum^{n}_{i=1}  \Lambda_i(1 - \Lambda_i)E(\bar{d}^{2}_{ij})
\label{eq_16}
\end{equation}
since the elements of $\Lambda_i = \Lambda(d_{i,j},\beta)$ are by definition non-stochastic in our case. Furthermore, \cite{arbia2015measurement} show that, under the hypothesis of uniform geo--masking expressed in Section \ref{sec:sec2}, we have:
\begin{equation}
E(\bar{d}^{2}_{ij}) = d^{2}_{ij} + \frac{\theta^*}{3}
\label{eq_17}
\end{equation}
(see also \cite{arbia2016spatial}). Expression \ref{eq_16}, therefore, can be re-written as:
\begin{equation}
Var(\hat{\bar{\beta}}) =  E  \left[  \sum^{n}_{i=1}  \Lambda_i(1 - \Lambda_i)  \left( d^{2}_{ij} + \frac{\theta^*}{3} \right)  \right] 
\label{eq_18}
\end{equation}
As a consequence, using Equation \ref{eq_19}, the efficiency loss due to geo-masking can be expressed as: 
\begin{equation}
EL = \frac{Var(\hat{\beta})}{Var(\hat{\bar{\beta}})} = \frac{E  \left[  \sum^{n}_{i=1}  \Lambda_i(1 - \Lambda_i)  \left( d^{2}_{ij} \right)  \right]}{ E  \left[  \sum^{n}_{i=1}  \Lambda_i(1 - \Lambda_i)  \left( d^{2}_{ij} + \frac{\theta^*}{3} \right)  \right]}.
\label{eq_20}
\end{equation}
Because the true distances $d_{ij}$ are deterministic, equation \ref{eq_20} clearly shows that the efficiency loss of the ML estimates of the parameter $\beta$ is an inverse function of the maximum displacement distance $\theta^{*}$.

\section{Discussion and conclusion}
\label{sec:sec5}

In this paper, we have examined the effects of measurement error introduced in a logistic model by random geo–masking of individual’s position, and distances are used as predictors. To analyse these effects, we started considering the case-study provided by \cite{berta2016association}, where the authors studied the determinants of the patients' choice for hospital admissions. Extending the classical results on the measurement error in a linear regression model, our Monte Carlo experiments on hospital choices showed that in a discrete choice model the higher is the distortion produced by the geo--masking, the higher will be the downward bias in absolute value towards zero of the coefficient associated to the distance. This effect can be seen as the discrete choice counter-part of the familiar "attenuation effect" very well known in the classical econometric literature on linear regression models \citep{greene2016econometric}. In particular, when a certain degree of locational error is introduced by arbitrarily allocating individuals in the centroid of a geographical partition (as it is customary to do in many empirical studies), according to intuition, the larger is the surface area of the geographical partition, the larger will be the downward bias. Furthermore, when data are intentionally displaced according to a random mechanism in order to protect confidentiality, we also observe a similar form of attenuation. The MC results show that this attenuation effect increases dramatically, with a bias that is observable as soon as we introduce even a small amounts of locational error. Furthermore, we provided a formal proof that the MLE of the logistic regression parameters lose efficiency when estimates are based on individuals whose position is intentionally geo--masked.\\
The results obtained in this paper could be used by the data producers to choose the optimal value of the parameters of geo--masking which preserves confidentiality, but does not destroys completely the statistical information. They can also be used to communicate to the practitioners the resulting level of attenuation and of efficiency loss they should expect from a logistic regression analysis performed using geo-masked datasets.

\bibliographystyle{chicago}
\bibliography{biblio}

\end{document}